\documentclass[twocolumn,showpacs,preprintnumbers,amsmath,amssymb]{revtex4}

\usepackage{graphicx}
\usepackage{dcolumn}
\usepackage{bm}
\usepackage{epsfig}

\begin{document}


\title{Fragility and compressibility at the glass transition}

\author{U. Buchenau}
\email{buchenau-juelich@t-online.de}
\author{A.Wischnewski}
\affiliation{%
Institut f\"ur Festk\"orperforschung, Forschungszentrum J\"ulich\\
Postfach 1913, D--52425 J\"ulich, Federal Republic of Germany
}%

\date{January 7, 2004; in revised form July 5, 2004}

\begin{abstract}
Isothermal compressibilities and Brillouin sound velocities from
the literature allow to separate the compressibility at the glass
transition into a high-frequency vibrational and a low-frequency
relaxational part. Their ratio shows the linear fragility relation
discovered by x-ray Brillouin scattering \cite{sco}, though the
data bend away from the line at higher fragilities. Using the
concept of constrained degrees of freedom, one can show that the
vibrational part follows the fragility-independent Lindemann
criterion; the fragility dependence seems to stem from the
relaxational part. The physical meaning of this finding is
discussed.
\end{abstract}


\maketitle

Very recently, Scopigno {\it et al} \cite{sco} compiled x-ray
Brillouin data in glasses, comparing the integrated Brillouin line
intensity to the intensity of the central line. They found a
linear relation between this ratio and the fragility
$m=\partial\log \eta/\partial(T_g/T)$, defined in terms of the
steep rise of the viscosity $\eta$ towards the glass temperature
$T_g$ with decreasing temperature in the supercooled liquid.

This striking result poses two questions. The first is a more
technical point: Does the momentum transfer range of the x-ray
Brillouin technique still reflect the long wavelength limit? The
second is more fundamental: Is the fragility related to the
Brillouin intensity or to the central line intensity? Scopigno
{\it et al} only demonstrate a linear relation to the {\it ratio}
of these two quantities. The present paper intends to address
these two questions by a comparison to literature data on the long
wavelength limit.

The total scattering of a supercooled liquid at low momentum
transfer is given by its isothermal compressibility $\chi_{T}$
\begin{equation}\label{sq0}
\lim _{Q\rightarrow 0} S(Q)=\rho \frac{kT}{M} \chi _{T},
\end{equation}
where $\rho$ is the density and $M$ is the average atomic mass.
Eq. (1) has been found to be valid for several molecular and
polymeric supercooled liquids \cite{fischdett} on the nm length
scale of the x-ray Brillouin technique.

The x-ray Brillouin experiment splits the total scattering $S(Q)$
into an apparently elastic central component $S_{IS}(Q)$ and two
Brillouin lines of summed intensity $S_{Brill}(Q)$. The
longitudinal sound velocity $v_{l\infty}$ at the Brillouin line
defines a high-frequency Brillouin compressibility
$\chi_{Brill}=1/\rho v_{l\infty}^{2}$. The ratio
$\alpha_{scatt}=S_{Brill}(Q)/S_{IS}(Q)$ at the glass temperature
$T_g$ reported by Scopigno et al \cite{sco} should equal the ratio
$\alpha_\chi$ between vibrational and relaxational compressibility
\begin{equation}\label{alfchi}
\alpha_\chi(T_g)=\frac{\chi_{Brill}(T_g)}{\chi_T(T_g)-\chi_{Brill}(T_g)}
\end{equation}

\begin{table}

\caption{Isothermal and Brillouin compressibility at the glass
transition.}

\begin{tabular}{|c|c|c|c|c|c|c|}
\hline
  substance & $T_g$ & $\rho$  & $v_{l\infty}$ & $v_{t\infty}$ & $\chi_{Brill}$ &
  $\chi_T$ \\
 & $K$ & $kg/m^3$  & $m/s$ & $m/s$  & $GPa^{-1}$ & $GPa^{-1}$ \\
\hline
  BeF$_2$   & 598$^a$  & 1900$^c$  & 4570$^c$ & & 0.0252 &   \\
  SiO$_2$   & 1450$^a$ & 2200$^d$  & 6480$^d$ & 3988$^d$ & 0.0108 &   \\
 B$_2$O$_3$ & 550$^e$  & 1792$^e$  & 3600$^f$ & 1933$^f$ & 0.0431 &
 0.39$^e$ \\
 PIB & 201$^b$  &  939$^g$  & 2994$^h$ &  & 0.119 &   \\
glycerol &187$^i$&1332$^i$&3583$^i$&1858$^j$&0.0586&0.287$^k$\\
   salol    &  218$^a$   & 1268$^l$ & 2382$^l$ & & 0.139 & \\
   1,4-PB   &  180$^a$   &  940$^g$  & 2500$^m$ & & 0.170 & \\
  PET &  342$^g$   & 1350$^g$  & 2309$^n$ & & 0.139 & 0.324$^o$ \\
 OTP     &  241$^a$   & 1124$^p$ & 2550$^q$ & & 0.137 & 0.39$^p$ \\
 Se   &  308$^a$   & 4262$^r$ & 2000$^s$ & & 0.0587 & 0.16$^r$ \\
 CKN &  343$^b$   & 2186$^e$  & 3190$^f$ & 1497$^f$ & 0.0450 & 0.132$^e$ \\
 PVAC  &  304$^e$  & 1186$^e$  & 2492$^t$ & & 0.136 & 0.498$^e$ \\
 BPA-PC &  418$^b$   & 1180$^u$  & 2176$^v$ & 938$^v$ & 0.179 & 0.511$^o$ \\
  PS   &  375$^b$   & 1028$^u$ & 2219$^w$ & & 0.198 & 0.558$^o$ \\
 PMMA  &  379$^b$   & 1161$^u$  & 2500$^x$ & 1278$^w$ & 0.138 & 0.473$^o$ \\
  PVC &  347$^x$   & 1372$^u$ & 2198$^x$ & & 0.151 & 0.385$^o$ \\
\hline
\end{tabular}

\

Abbreviations: PIB=polyisobutylene; PB=polybutadiene;
PET=polyethylenteraphtalate; OTP=orthoterphenyl;
CKN=K$_3$Ca$_2$(NO$_3$)$_7$; PVAC=polyvinylacetate;
BPA-PC=polycarbonate; PS=polystyrene; PMMA=polymethylmethacrylate;
PVC=polyvinylchloride.

\

$^a$ref. \cite{sco}; $^b$ref. \cite{bohmer}; $^c$ref. \cite{sco2};
$^d$ref. \cite{bucaro}; $^e$ref. \cite{gupta}; $^f$ref.
\cite{grims}; $^g$ref. \cite{krevelen}; $^h$ref. \cite{sokolov};
$^i$ref. \cite{comez}; $^j$ref. \cite{fio2}$^k$ref.
\cite{gilchrist}; $^l$ref. \cite{dreyfus}; $^m$ref. \cite{fio};
$^n$ref. \cite{pattpet}; $^o$ref. \cite{schwarzlchi}; $^p$ref.
\cite{naoki}; $^q$ref. \cite{monaco}; $^r$ref. \cite{berg};
$^s$ref. \cite{ruocco}; $^t$ref. \cite{jimenez}; $^u$ref.
\cite{schwarzlrho}; $^v$ref. \cite{pattbpapc}; $^w$ref.
\cite{kruegerps}; $^x$ref. \cite{pentecost};

\end{table}

Table I compiles literature data of the isothermal compressibility
and the Brillouin sound velocity. Most of the Brillouin sound
velocities in Table I were obtained by light scattering; at the
glass transition, light and x-ray scattering sound velocities
still agree \cite{comez,fio,monaco}.

\begin{figure}[b]
\hspace{-0cm} \vspace{0cm} \epsfig{file=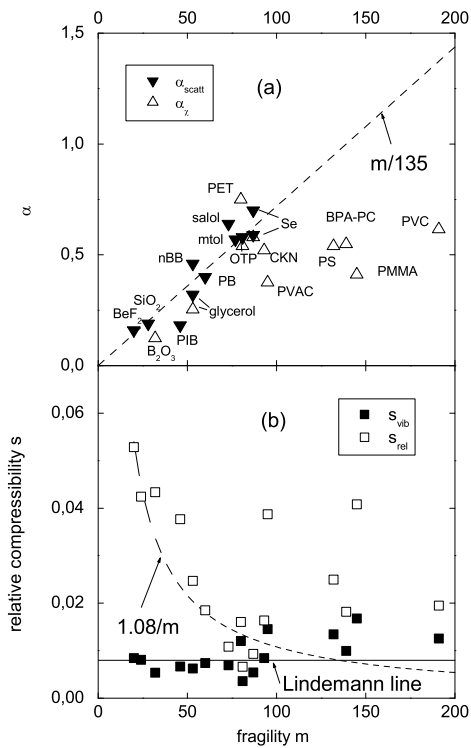,width=7
cm,angle=0} \vspace{0cm} \caption{(a) Comparison of
$\alpha$-values from x-ray Brillouin scattering and from the
long-wavelength limit. (b) Fragility dependence of normalized
vibrational and relaxational compressibilities.}
\end{figure}

As shown in Fig. 1 (a), the general tendency is the same and the
three common cases glycerol, OTP and Se agree reasonably well in
both sets of data. The literature data go to much higher fragility
and thus reveal a bending away from the postulated line: at higher
fragilities, the compressibility ratio becomes
fragility-independent. But even so, the answer to our first
question is clear: the relation discovered by x-ray Brillouin
scattering \cite{sco} is a true property of the long-wavelength
limit.

In order to answer the second question, namely whether the
Brillouin line or the central line provides the relation to the
fragility, one needs to compare the vibrational or relaxational
compressibility of different glass formers. This requires an
appropriate normalization of the compressibility. It is reasonable
to define a dimensionless ratio between the thermal energy at the
glass transition and the vibrational and relaxational compression
energy, respectively, by
\begin{equation}\label{cnorm}
s_{vib}=\frac{k_BT_g\chi_{Brill}}{v}\ \ {\rm and}\
s_{rel}=\frac{k_BT_g(\chi_T-\chi_{Brill})}{v},
\end{equation}
where $v$ is an appropriate microscopic volume.

{\it A priori}, one would choose for $v$ the atomic volume.
However, this choice is not justified, because glass formers are
complex solids, with a mixture of strong and weak nearest-neighbor
bonds \cite{phillips,thorpe}. For instance, a polymer owes its
material properties to a mixture of covalent and van-der-Waals
bonds, very different in strength. Therefore one must take this
difference in bonding strength into account.

\begin{table}

\caption{Ratio of vibrational and relaxational compressibility at
the glass transition.}

\begin{tabular}{|c|c|c|c|c|c|c|}
\hline
 substance & $\alpha_\chi$ & $\alpha_{scatt}$ & $f_s$ & $M$ & $B/B_0$ & $m$ \\
 &   &  &  & 10$^{-27}kg$ & \\
\hline
 BeF$_2$ &  & 0.16$^a$ & 5/9 & 26.0 & & 20$^a$ \\
 SiO$_2$ &  & 0.191$^a$ & 5/9 & 33.25 & & 24$^b$ \\
 B$_2$O$_3$ & 0.124 &   & 1/5 & 21.92 & 5.6 & 32$^b$ \\
 PIB &  & 0.182$^y$ & 1/6 & 7.75 & & 46$^b$ \\
 glycerol & 0.253   & 0.32$^a$ & 1/3 & 10.91 & 3.1 & 53$^a$ \\
 salol  &  & 0.64$^a$ & 7/39 & 13.66 & & 73$^b$ \\
 1,4-PB    &  & 0.40$^a$ & 1/6 & 8.96 &  & 60$^a$ \\
 PET  &  0.751  &  & 13/66 & 14.49 & & 80$^z$ \\
 OTP &   0.540  & 0.58$^a$ & 1/12 & 11.93 & & 81$^b$ \\
 Se &  0.579  & 0.7$^a$ & 2/3 & 131.1 & & 87$^b$ \\
 CKN  & 0.516  &  & 19/33 & 31.76 & 2.1 & 93$^b$ \\
 PVAC &  0.375  &  & 1/4 & 11.9 & & 95$^b$ \\
 BPA-PC & 0.539   &  & 14/99 & 12.78 & 2.1 & 132$^b$ \\
 PS  & 0.548   &  & 5/48 & 10.97 & & 139$^b$ \\
 PMMA   & 0.411   &  & 2/9 & 11.07 & 2.2 & 145$^b$ \\
 PVC    & 0.644  &  & 4/21 & 15.05 & & 191$^b$ \\
\hline
\end{tabular}

$^y$ref. \cite{farago}; $^z$ref. \cite{saiter}; other see Table I.

\end{table}

Thus we define $v=v_{at}/f_s$, where $v_{at}$ is the atomic volume
and $f_s$ is the fraction of soft degrees of freedom in the
substance. The $f_s$-values in Table II were calculated assuming
the stretching of all covalent bonds (including the Be-F bond) as
well as the bond bending at boron, nitrogen and carbon to be hard.
All other degrees of freedom were considered to be soft. The
resulting $s_{vib}$ and $s_{rel}$ values are shown in Fig. 1 (b).
Though the scatter of points is even worse than in Fig. 1 (a), one
observes that the fragility rise is only weakly correlated with
the vibrational softening, but strongly with a decrease of the
relaxational compressibility.

The first part of this result is consistent with empirical
knowledge: The glass temperature tends to be about a factor of 0.6
smaller than the melting temperature $T_m$, which in turn follows
the empirical Lindemann criterion \cite{lindemann}. The Lindemann
criterion states a mean square vibrational displacement of the
atoms in the crystal of 10 percent of the nearest neighbour
distance at the melting point. On the basis of a Debye model and a
constant ratio of transverse and longitudinal sound velocity, one
then expects the same vibrational compressibility in all glass
formers at $T_g$, independent of the fragility $m$.

To quantify this Lindemann expectation, we assume a nearest
neighbour distance $d\approx v_{at}^{1/3}$ and an average ratio of
longitudinal to transverse sound velocity
$v_{l\infty}/v_{t\infty}\approx1.8$. The Lindemann criterion in
the form improved by Gilvarry \cite{gilvarry} reads
\begin{equation}\label{linde}
<u^2>(T_m)=\frac{3k_BT_m}{M\omega_D^2}\equiv (0.083d)^2.
\end{equation}
Here $<u^2>$ is the mean square displacement in {\it one}
direction and $\omega_D$ is the Debye frequency
\begin{equation}\label{omd}
\omega_D^3=\frac{18\pi^2}{v_{at}(1/v_l^3+2/v_t^3)}.
\end{equation}
With the above assumptions (including $T_g\approx 0.6T_m$), one
finds
\begin{equation}\label{lindeline}
k_BT_g\approx 0.008\frac{v_{at}}{\chi_{Brill}}.
\end{equation}
This is the Lindemann line in Fig. 1 (b), which is in reasonable
agreement with the data points for the vibrational
compressibility. Taking this Lindemann compressibility and the
linear relation $\alpha=m/135$ of Scopigno {\it et al} (the dashed
line in Fig. 1 (a)) , one gets $s_{rel}=1.08/m$, the dashed line
in Fig. 1 (b). We observe that this relation is only followed in
the lower half of the fragility region, consistent with the
bending away observed in Fig. 1 (a).

The temperature dependence of the ratio between relaxational and
vibrational compressibility $1/\alpha_\chi$ is shown in Fig. 2 for
three of the glass formers of Table I. There is no strong decrease
of the ratio towards $T_g$, so the relaxational compressibility is
neither proportional to the free volume nor to the excess entropy
of the glass former \cite{jaeckle}. In the first case, it should
extrapolate to zero at the Vogel-Fulcher temperature, in the
second at the Kauzmann temperature (these two temperatures tend to
lie close to each other \cite{ang1,ang2,ang3}).

To clarify the physical meaning of the compressibility ratio, it
is useful to rewrite it in terms of elastic moduli. Let $B$ and
$G$ be the infinite-frequency bulk and shear modulus,
respectively. $\chi_T=1/B_0$, where $B_0$ is the zero frequency
bulk modulus (the zero frequency shear modulus is zero above
$T_g$). $1/\chi_{Brill}=B+4G/3\approx 1.7B$, where we used again
the approximation $v_{l\infty}/v_{t\infty}\approx 1.8$ mentioned
in the derivation of the Lindemann line. Thus
\begin{equation}\label{elast}
1/\alpha_\chi\approx1.7\frac{B}{B_0}-1,
\end{equation}
so the ratio between relaxational and vibrational compressibility
is a measure for the ratio between high frequency and low
frequency bulk modulus. For strong glasses, this is high, for
fragile glasses, it is low. In fact, for the five glasses where we
know not only the isothermal compressibility, but also both the
longitudinal and the transverse Brillouin sound velocity (see
Table I), the calculated ratio $B/B_0$ in Table II decreases with
increasing fragility (in those cases, it is possible to determine
$B/B_0$ directly from experiment without any approximation).

The question is: What determines the ratio $B/B_0$ between
long-time and short-time bulk modulus at the glass transition?
This question can be translated into another question: What
happens to the bulk modulus in the relaxation processes which
bring the shear modulus down to zero? There are two extremes: (i)
The bulk modulus is also brought down to zero (ii) The bulk
modulus is not affected at all. In the first case, $B/B_0$ is
infinite and $\alpha_\chi=0$, in the second case $B/B_0=1$ and
$\alpha_\chi\approx 1.43$. However, these extremes are never
reached; in Table II, $\alpha_\chi$ ranges from 0.124 to 0.751, so
$B/B_0$ ranges from 1.35 to 5.6.

The consideration helps to understand the weakness of the
temperature dependence in Fig. 2: $B/B_0$ changes only slowly with
temperature (if it changes at all). One also understands the
physical meaning of the relation found by Scopigno {\it et al}
\cite{sco}: In strong glass formers, $B/B_0$ is large, in fragile
ones small. Strong glass formers show strong relaxational density
fluctuations on the scale of their vibrational compressibility, a
factor of three to four stronger than fragile ones.

\begin{figure}[b]
\hspace{-0cm} \vspace{0cm} \epsfig{file=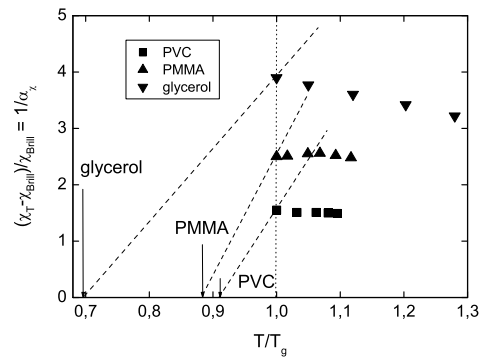,width=7
cm,angle=0} \vspace{0cm} \caption{Ratio of relaxational to
vibrational compressibility for PVC, PMMA and glycerol above $T_g$
(the references are the same as in Table I). The arrows mark the
Vogel-Fulcher temperatures of the three glass formers, the dashed
lines show the free volume expectation.}
\end{figure}

\begin{acknowledgments}
Helpful discussions with Giancarlo Ruocco and Reiner Zorn are
gratefully acknowledged.
\end{acknowledgments}


\begin{thebibliography}{99}
\bibitem{sco} T. Scopigno, G. Ruocco, F. Sette and G. Monaco, Science
{\bf 302}, 849 (2003)
\bibitem{fischdett} E. W. Fischer and M. Dettenmaier, J. Non-Cryst.
Solids \textbf{31}, 181 (1978) and further references therein
\bibitem{bohmer} R. B\"ohmer, K. L. Ngai, C. A. Angell, D. J.
Plazek, J. Chem. Phys. \textbf{99}, 4201 (1993)
\bibitem{sco2} T. Scopigno, S. N. Yannopoulos, D. Th.
Kastrissios, G. Monaco, E. Pontecorvo, G. Ruocco and F. Sette, J.
Chem. Phys. \textbf{118}, 311 (2003)
\bibitem{bucaro} J. A. Bucaro and H. D.
Dardy, J. Appl. Phys. \textbf{45}, 5324 (1974)
\bibitem{gupta} P. K.
Gupta, C. T. Moynihan, J. Chem. Phys. \textbf{65}, 4136 (1976)
\bibitem{grims} M. Grimsditch, L. M. Torell, in
\textit{Dynamics of Disordered Materials}, Springer Proceedings in
Physics \textbf{37}, ed. D. Richter, A.J. Dianoux, W. Petry and J.
Texeira, Springer Verlag Berlin 1989, p.196; L. M. Torell and R.
Aronsson, J. Chem. Phys. \textbf{78}, 1121 (1983)
\bibitem{krevelen} D. W. van
Krevelen, \textit{Properties of Polymers}, (Elsevier, Amsterdam
1976), p. 70
\bibitem{sokolov} A. P. Sokolov (private communication)
\bibitem{comez} L. Comez, D. Fioretto, F.
Scarponi, G. Monaco, cond-mat/0305348
\bibitem{fio2} F. Scarponi, L. Comez, D. Fioretto, L. Palmieri,
Phil. Mag. 2004 (in press)
\bibitem{gilchrist} A. Gilchrist, J. E. Earley, R.
H. Cole, J. Chem. Phys. \textbf{26}, 196 (1957)
\bibitem{dreyfus} C. Dreyfus, M. J. Lebon, H. Z.
Cummins, J. Toulouse, B. Bonello and R. M. Pick, Phys. Rev. Lett.
\textbf{69}, 3666 (1992)
\bibitem{fio} D. Fioretto, U.
Buchenau, L. Comez, A. Sokolov, C. Masciovecchio, A. Mermet, G.
Ruocco, F. Sette, L. Willner, B. Frick, D. Richter and L. Verdini,
Phys. Rev. E \textbf{59}, 4470 (1999)
\bibitem{pattpet} G. D. Patterson, J. Polym. Sci.:
Polym. Phys. \textbf{14}, 1909 (1976)
\bibitem{schwarzlchi} F. R. Schwarzl,
\textit{Polymermechanik,} (Springer, Berlin 1990), Tabelle 12.1
\bibitem{naoki}
M. Naoki and S. Koeda, J. Phys. Chem. \textbf{93}, 948 (1989)
\bibitem{monaco} G. Monaco, D. Fioretto, L. Comez and
G. Ruocco, Phys. Rev. E \textbf{63}, 061502 (2001)
\bibitem{berg} J. I. Berg and
R. Simha, J. Non-Cryst. Solids \textbf{22}, 1 (1976)
\bibitem{ruocco} G. Ruocco, private communication
\bibitem{jimenez} J. K. Kr\"{u}ger, K. P.
Bohn, R. Jimenez and J. Schreiber, Colloid Polym. Sci.
\textbf{274}, 490 (1996)
\bibitem{schwarzlrho} F. R. Schwarzl,
\textit{Polymermechanik,} (Springer, Berlin 1990), Abb. 6.12
\bibitem{pattbpapc} G. D. Patterson, J. Polym.
Sci: Polym. Phys. Ed. \textbf{14}, 741 (1976)
\bibitem{kruegerps} H.
Kr\"{u}ger, in \textit{Optical Techniques to Characterize Polymer
Systems}, Studies in Polymer Science 5, ed H. B\"{a}ssler
(Elsevier, Amsterdam 1989), p. 491
\bibitem{pentecost} D. A. Jackson, H. T. A.
Pentecost and J. G. Powles, Mol. Phys. \textbf{23}, 425 (1972)
\bibitem{phillips} J. C. Phillips, J. Noncryst. Solids {\bf 34},
153 (1979)
\bibitem{thorpe} M. F. Thorpe, J. Noncryst. Solids {\bf 57}, 355
(1983)
\bibitem{farago} B. Farago, A. Arbe, J.
Colmenero, R. Faust, U. Buchenau and D. Richter, Phys. Rev. E
\textbf{65}, 051803 (2002) obtain $S_{IS}(Q)=0.22$ in a neutron
experiment; $\alpha_{scatt}$ follows from eqs. (1) and (2)
\bibitem{saiter} J. M. Saiter, E. Dargent, M. Kattan, C. Cabot and J. Grenet,
Polymer \textbf{44}, 3995 (2003)
\bibitem{lindemann} F. A. Lindemann, Phys. Z. {\bf 11}, 609 (1910)
\bibitem{gilvarry} J. J. Gilvarry, Phys. Rev. {\bf 102}, 308
(1956)
\bibitem{jaeckle} J. J\"ackle, Rep. Prog. Phys. {\bf 49}, 171
(1986)
\bibitem{ang1} C. A. Angell, J. Non-Cryst. Solids {\bf 131-133},
13 (1991)
\bibitem{ang2} C. A. Angell, in {\it Proc. Int. School of
Physics, "Enrico Fermi" Course CXXXIV} edited by F. Mallamace and
H. E. Stanley, IOS Press Amsterdam, 1997, p. 571.
\bibitem{ang3} C. A. Angell, J. Res. NIST 102, 171
(1997)
\end{thebibliography}
\end{document}